\documentclass[letter,11pt]{article}
\usepackage{graphicx}
\usepackage{amsmath}
\usepackage{pstricks}
\usepackage{epsfig}

\newcommand{\SLACPubNumber} {10065}
\newcommand{\LANLNumber} {0307045}

\def\Bbar  {\kern 0.18em\overline{\kern -0.18em B}{}}
\def\Bz    {\ensuremath{B^0}}
\def\Bzb   {\ensuremath{\Bbar^0}}
\def\BzBzb {\ensuremath{B^0 {\kern -0.16em \Bzb}}}

\def\qqbar {\ensuremath{q\overline q}}

\def\Dt    {\ensuremath{{\rm \Delta}t}}
\def\Dm    {\ensuremath{{\rm \Delta}m}}
\def\to    {\ensuremath{\rightarrow}}
\newcommand{\jprl}      [1]  {{Phys.\ Rev.\ Lett.\ {\bf #1}}} 
\newcommand{\progtp}    [1]  {{Prog.\ Th.\ Phys.\ {\bf #1}}}
\newcommand{\mpl}       [1]  {{Mod.\ Phys.\ Lett.\ {\bf #1}}}
\newcommand{\np}        [1]  {{Nucl.\ Phys.\ {\bf #1}}}
\newcommand{\pr}        [1]  {{Phys.\ Rev.\ {\bf #1}}}
\newcommand{\VudVub}{V_{ud}^{}V_{ub}^*}
\newcommand{\VcdVcb}{V_{cd}^{}V_{cb}^*}
\newcommand{\VtdVtb}{V_{td}^{}V_{tb}^*}
\newcommand{\Btag}{B_{\mathrm{tag}}}

\newcommand{\epstag}{\epsilon_{\mathrm{tag}}}
\newcommand{\epseff}{\epsilon_{\mathrm{eff}}}
\newcommand{\Bflav}{B_{\mathrm{flav}}}
\newcommand{\Amix}{A_{\mathrm{mix}}}
\newcommand{\Nmixed}{N_{\mathrm{mixed}}}
\newcommand{\Nunmixed}{N_{\mathrm{unmixed}}}

\long\def\inst#1{\par\nobreak\kern 4pt\nobreak
    {\it #1}\par\vskip 10pt plus 3pt minus 3pt}

\begin{document}
{\pagestyle{empty}

\begin{flushright}
SLAC-PUB-\SLACPubNumber \\
hep-ex/\LANLNumber \\
July, 2003 \\
\end{flushright}

\par\vskip 4cm

\begin{center}
\Large \bf 
Recent measurements of $CP$ violation at the B factories 
\end{center}
\bigskip

\begin{center}
\large 
Gabriella Sciolla\\
 Massachusetts Institute of Technology\\
 77 Massachusetts Avenue, 
Cambridge, MA 02139-4307, USA\\
sciolla@mit.edu
\end{center}
\bigskip \bigskip

\begin{center}
\large \bf Abstract
\end{center}
Recent measurements of time dependent $CP$ asymmetries at the $B$ factories 
have led to substantial progress in our understanding of $CP$ violation. 
In this article, I review some of these experimental results and  
discuss their implications in the Standard Model and 
their sensitivity to New Physics. 

\vfill
\begin{center}
To be published in  Modern Physics Letters A. 
\end{center}

\vspace{1.0cm}
\begin{center}
\hrule\vspace{0.1cm}
Work supported in part by Department of Energy contract DE-FC02-94ER40818.
\end{center}

\newpage

\section{Introduction}	

$CP$ violation is one of the most intriguing and least understood topics in particle physics.
According to Sakharov\cite{sakarov}, $CP$ violation is a necessary condition  
to explain how equal amounts of matter and anti-matter created in the Big Bang 
may have evolved into a matter-dominated universe. 
Thus $CP$ violation is a requisite to our own existence. 

$CP$ violation  was discovered\cite{fitch}   in 1964
in the decays of $K_L$ mesons into two pion final states. 
A simple and elegant explanation of this effect in the context of the Standard Model 
was proposed by Kobayashi and Maskawa\cite{KM} in 1972. In the Kobayashi-Maskawa mechanism, 
$CP$ violation originates from a single complex phase in the  
mixing matrix between the three quark families.  
This mechanism, however, does not allow for enough $CP$ violation to explain the 
matter-dominated universe\cite{prd-ref6}.
Understanding the mechanism that governs this phenomenon, and eventually uncovering  
its origin, remains one of the central questions in modern physics. 

Many precise measurements of $CP$ violation have been made in the study of decays of 
neutral kaons\cite{pdg}. However, due to hadronic uncertainties, these measurements 
do not pose significant constraints on the parameters of the theory. 
In contrast, large $CP$ violation effects  essentially free of hadronic uncertainties 
are expected in final states of $B$ meson decays.

Two asymmetric $B$ factories, PEP-II at SLAC and KEKB at KEK, were designed 
for studying $CP$ violation in the $B$ system
as their main goal. Since the beginning of data taking in 1999, the $B$ factories 
recorded unprecedented samples of $B$ mesons and started a new era in the study of 
$CP$ violation and $B$ physics. 

In this article,
I review some of the results  on $CP$ violation  obtained so far at the $B$ factories 
and discuss how these measurements constrain the Standard Model, 
and how they can be used to probe New Physics.

\section{$CP$ violation in the Standard Model}	
In the Standard Model, $CP$ violation originates from a complex phase in the 
quark-mixing matrix, the Cabibbo-Kobayashi-Maskawa (CKM) matrix. 
Following Wolfenstein's notation\cite{wolf}, the CKM matrix can be expressed in 
terms of four real parameters $\lambda$, $A$, $\rho$ and $\eta$ as  
%
\[
V = 
\begin{pmatrix}
V_{ud} & V_{us} & V_{ub} \\
V_{cd} & V_{cs} & V_{cb} \\
V_{td} & V_{ts} & V_{tb}
\end{pmatrix}
= 
\begin{pmatrix}
1-\lambda^2/2            &   \lambda       & A\lambda^3(\rho-i\eta) \\
-\lambda                 &   1-\lambda^2/2        & A\lambda^2 \\
A\lambda^3(1-\rho-i\eta) &    -A\lambda^2      & 1
\end{pmatrix}
+ O(\lambda^4).
\]
The coefficient $\lambda$ is the sine of the Cabibbo angle, which is 
well measured from studies of $K^+\to \pi^0 l^+ \nu$ decays.  
The coefficient $A$ is related to the CKM matrix element $V_{cb}$ and can be determined 
from the study of semileptonic $B$ decays such as $B\rightarrow D^* l \nu$. 
The parameters $\rho$ and $\eta$ are related to $CP$ violation in the $B$ system. 

The CKM matrix is unitary, i.e. $V^\dag V=1$. This implies six equations that relate the 
elements of the matrix. One of these equations, 
$\VudVub+\VcdVcb+\VtdVtb=0$, is 
particularly useful for studies of $CP$ violation in the $B$ system.
Dividing all the elements of the sum by $\VcdVcb$, we 
obtain the ``Unitarity Triangle'' (UT)  shown in figure \ref{fig:ut} in the ($\rho$,$\eta$) plane. 
\vspace{1cm}
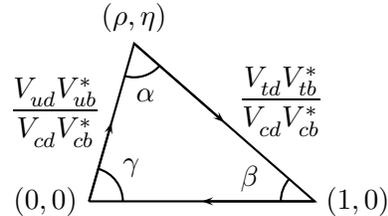
\begin{figure}[th]
  \begin{center}
    \psset{unit=3cm}
    \begin{pspicture}(-0.3,-0.05)(1.3,0.75)
      \psline(0,0)(1,0)(0.2,0.7)(0,0)
      \psline{->}(0,0)(0.1,0.35)
      \psline{->}(0.2,0.7)(0.6,0.35)
      \psline{->}(1,0)(0.5,0)
      \psarc(0,0){0.15}{0}{74}
      \psarc(0.2,0.7){0.15}{254}{319}
      \psarc(1,0){0.15}{139}{180}
      \rput[r](-0.05,0){$(0,0)$}
      \rput[l](1.05,0){$(1,0)$}
      \rput[b](0.2,0.75){$(\rho,\eta)$}
      \rput[r](0.05,0.40){\parbox{1.2cm}{$$\frac{\VudVub}{\VcdVcb}$$}}
      \rput[l](0.65,0.45){\parbox{1.2cm}{$$\frac{\VtdVtb}{\VcdVcb}$$}}
      \rput[t](0.25,0.50){$\alpha$}
      \rput[r](0.75,0.10){$\beta$}
      \rput[l](0.15,0.15){$\gamma$}
    \end{pspicture}
  \end{center}
  \caption{The Unitarity Triangle.}
  \label{fig:ut}
\end{figure}

Because the lengths of the sides of the triangle are of the same order, 
the angles can be large ($O(1)$), 
leading to potentially large $CP$-violating asymmetries 
from phases between CKM matrix elements. This means that for this triangle one 
can experimentally measure both sides and angles. 
These measurements over-constrain the triangle, thus allowing for consistency checks 
of the $CP$ sector of the Standard Model. 

The side $\VudVub/\VcdVcb$ is constrained by measurements of 
$|V_{ub}/V_{cb}|$ from studies of  $b\to ul\nu$ and $b\rightarrow c l \nu$ transitions. 
The side $\VtdVtb/\VcdVcb$ is constrained by measurements of 
 \Bz\ and $B_s^0$ mixing frequencies. 
Another constraint on the apex of the Unitarity Triangle comes from the measurement of 
the $CP$ violation parameter $|\epsilon_K|$ in the kaon system. 
The measurements of the angles come from $CP$ violation studies in $B$ decays. 

A comparison between the measurement of the angles and the position of the apex of the 
UT in the ($\rho$,$\eta$) plane allows a quantitative test of the $CP$ sector 
of the Standard Model.
Studies of $CP$ violation in the $B$ system can also be used as a probe for 
New Physics 
that could enhance the role of box or penguin diagrams relative to the tree diagrams,  
due to the contribution of new (virtual) particles participating in the loops.

\section{$CP$ violation in $B^0$ decays }	
We define \Bz\ $(=\overline{b} d)$ and \Bzb\ $(=b\overline{d})$ the neutral $B$ meson flavor eigenstates,
 and $B_H$ and $B_L$ the eigenstates of the
Hamiltonian, with definite mass and lifetime. The mass eigenstates can be expressed in terms of the 
flavor eigenstates: 
\[
|B_L\rangle = p |\Bz\rangle+ q |\Bzb\rangle 
\quad\text{and}\quad
|B_H\rangle = p |\Bz\rangle - q |\Bzb\rangle,
\]
where $p$ and $q$ are complex coefficients that satisfy the condition $ |q|^2 + |p|^2 = 1$.

The time evolution of a pure \Bz\ or \Bzb\ state at time $t=0$,
is given by:
\begin{equation}\begin{split}
|\Bz(t)\rangle
& = e^{-im_Bt} e^{-\Gamma t/2} \left \{ \cos \frac{ \Delta m t}{2} |\Bz\rangle 
                             -i  \frac{q}{p} \sin \frac{ \Delta m t}{2} |\Bzb\rangle \right \},\\
|\Bzb(t)\rangle
& = e^{-im_Bt} e^{-\Gamma t/2} \left \{ \cos \frac{ \Delta m t}{2} |\Bzb\rangle 
                             -i  \frac{p}{q} \sin \frac{ \Delta m t}{2} |\Bz\rangle \right \},
\label{timeEvol}
\end{split}\end{equation}
where $m_B=(m_H+m_L)/2$ and $\Gamma=\Gamma_H \sim \Gamma_L $.

Let's consider decays of \Bz\ and \Bzb\ mesons into a final state that is a $CP$ eigenstate, $f_{CP}$, and 
define the two decay amplitudes as 
\[
A_f=\langle f_{CP}|H|B^0\rangle 
\quad\text{and}\quad 
\overline{A}_f=\langle f_{CP}|H|\overline{B}^0\rangle. 
\]

The probability for a  \Bz\ or a  \Bzb\ to decay in the final state $f_{CP}$ 
at the time $t$ will be proportional to the square of the time dependent
amplitudes 
$A_f(t)=\langle f_{CP}|H|B^0(t)\rangle$  and 
$\overline{A}_f(t)= \langle f_{CP}|H|\overline{B}^0(t)\rangle$:  
\begin{equation}\begin{split}
N(\Bz(t)\to f_{CP}) &\propto  e^{-\Gamma t} \left \{
1 -  \frac{2\text{Im}\lambda_f }{1+|\lambda_f |^2} \sin (\Dm t) + \frac{1-|\lambda_f|^2}{1+|\lambda_f|^2}\cos (\Dm t) 
\right  \}, \\
N(\Bzb(t)\to f_{CP}) &\propto e^{-\Gamma t} \left \{
1 + \frac{2\text{Im}\lambda_f }{1+|\lambda_f |^2} \sin (\Dm t) - \frac{1-|\lambda_f|^2}{1+|\lambda_f|^2}\cos(\Dm t) 
\right  \},  
\label{mixProbab}
\end{split}\end{equation}
where
\begin{equation} 
\lambda_f \equiv  \eta_f \frac{q}{p}  \frac{\overline{A}_f}{A_f},
\label{lambda}
\end{equation}

with $\eta_f$ being the $CP$ eigenvalue of the final state $f_{CP}$,
and \Dm\  the mass difference between the mass eigenstates
$B_H$ and $B_L$. 

We define the time dependent $CP$ asymmetry as:   
\begin{equation}
		A_{CP}(t) \equiv \frac{N(\Bzb(t)\to f_{CP}) - N(\Bz(t)\to f_{CP})} {N(\Bzb(t)\to f_{CP}) + N(\Bz(t)\to f_{CP})}.
\label{acpt}
\end{equation}
Substituting (\ref{mixProbab}) into the definition (\ref{acpt}), it follows that 
\begin{equation}
		A_{CP}(t) =  S_f \sin(\Delta m t) - C_f \cos(\Delta m t), 
\label{acpt2}
\end{equation}
where 
\begin{equation}
C_f = \frac{1-|\lambda _f|^2}{1+|\lambda _f |^2}
\quad\text{and}\quad
S_f = \frac{2 \text{Im} \lambda_f}{1+|\lambda _f|^2}. 
\label{acpt3}
\end{equation}

For \Bz\ decays, $|q/p| \sim 1$.
If only one diagram contributes to the final state,  
as in the case of \Bz \to $J/\psi K_S$,
$|\overline{A}_f/A_f|=1$, thus $|\lambda_f|=1$.
This simplifies equation (\ref{acpt2})
leaving the time dependent $CP$ asymmetry with only a sine component having  amplitude  $\text{Im}(\lambda_f )$:
\begin{equation}
A_{CP}(t) = \text{Im}(\lambda_f ) \sin(\Delta m t). 
\label{acpt4}
\end{equation}
For these final states, $\text{Im}(\lambda_f )$ is directly and simply related to the angles of
the UT triangle.

If more than one diagram contributes to the final state $f_{CP}$,
then $A_{CP}(t)$ maintains both the sine and cosine components.
The coefficient $S_f$ is still related to the 
the angles of the UT, while $C_f$ measures direct $CP$ violation.

\section{The experimental apparatus  }	

The two asymmetric $B$ factories,  
PEP-II\cite{pepii}  at SLAC and KEKB\cite{kekb} at KEK,
have similar designs. 
They are both $e^+e^-$ colliders operating at 
a center of mass energy of $\sqrt{s}=10.58$ GeV,
the mass of the $\Upsilon(4S)$ resonance. 
This resonance decays exclusively to \Bz\Bzb\ and $B^+B^-$ pairs,
providing  ideal conditions for the study of $B$ meson decays. 

PEP-II collides 9.0 GeV electron and 3.1 GeV positron beams head-on,
producing $\Upsilon (4S)$ with a Lorentz boost of $\beta\gamma$ = 0.56 along the electron beam axis. 
KEKB, on the other hand,  collides 8.0 GeV electron and 3.5 GeV positron beams at a small 
($\pm$11 mrad) crossing angle, producing $\Upsilon (4S)$ with  $\beta\gamma=0.425$. 
These boosts are  essential to $CP$ violation studies because they allow
the experiments to separate 
the decay vertices of the two $B$ mesons, and thus to  measure the  time dependence of their 
decay rates. 

Each machine hosts a large solid angle general purpose detector: 
BaBar\cite{babar} at PEP-II and Belle\cite{belle} at KEKB. 
Both detectors are equipped with silicon vertex detectors, cylindrical drift chambers, 
Cherenkov detectors for particle identification, crystal calorimeters and muon detection systems. 
The magnetic field is provided by 1.5 T superconducting solenoidal coils.

Both facilities have been running very successfully since 1999, reaching or 
surpassing their design luminosities: PEP-II doubled the design goal and
achieved 6$\times$10$^{33}$\,cm$^{-2}$s$^{-1}$,  
while  KEKB has reached the design luminosity of 1$\times$10$^{34}$\,cm$^{-2}$s$^{-1}$.
The integrated luminosities delivered  to date by the two machines are  
136\,fb$^{-1}$ for PEP-II and 155\,fb$^{-1}$ for KEKB. 
Of these, about 82 fb$^{-1}$ and 78 fb$^{-1}$ have been 
analyzed by the BaBar and Belle Collaborations, respectively,  
to produce the results discussed in this article.

\section{The measurement of $A_{CP}(t)$ at the $B$ factories }	

At the $B$ factories, $CP$ violation is studied through the measurement of the time dependent $CP$ 
asymmetry, $A_{CP}(t)$, defined in (\ref{acpt}). 
The measurement utilizes those decays of the $\Upsilon (4S)$ into two neutral $B$ mesons, 
of which one ($B_{CP}$) can be completely  reconstructed  into a $CP$ eigenstate, 
while the decay products of the other ($\Btag$) are measured in an attempt
 to infer its flavor at decay time. 

Due to the intrinsic spin of the $\Upsilon(4S)$, the \Bz\Bzb\ pair is produced in a coherent $L=1$ state. 
After production, each meson will evolve in time as given in equation (\ref{timeEvol}). 
Because the net quantum numbers of the system are conserved, the two mesons evolve in phase 
until one of them decays. When the first $B$ meson decays into a flavor eigenstate, 
the other $B$ is dictated to be of opposite flavor at that same instant. 
For this reason the time that appears in (\ref{acpt}) can be considered as 
the difference in time between the two decays (\Dt ). 

A schematic view of a typical event used for $CP$ analysis is given in figure \ref{AnalSketch}.  
The logical steps of the analysis are discussed in the following subsections.  
More detailed information about the analysis techniques can be found in 
references~\cite{babarPRD} and~\cite{bellePRD}. 

\begin{figure}[th]
\begin{center}
\includegraphics[width=0.65\linewidth]{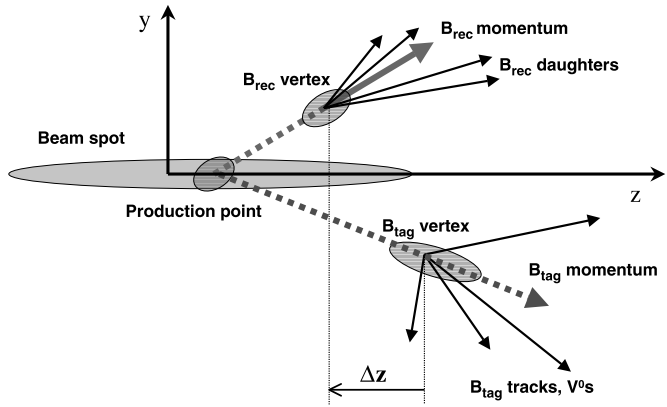}
\end{center}
\caption{Schematic view in the $y$-$z$ plane of a $\Upsilon(4S)\to\Bz\Bzb$ decay. Note that the 
scale in the $y$ direction is substantially magnified compared to that in the $z$ direction. }
\label{AnalSketch}
\end{figure}

\subsection{Reconstruction of $CP$ eigenstates}
 
Abundant and pure exclusive reconstruction of $CP$ final states is the first step 
to a successful measurement. 
Since the branching fractions of  experimentally accessible states are small 
(between 10$^{-6}$ and 10$^{-4}$) it is crucial to use as many $CP$ modes as possible to enhance the 
statistical significance of the measurement. In order  to preserve the interpretation 
of the measurements, only decays  characterized by the same Feynman  diagrams 
are combined. 

\subsection{Flavor tagging}

The flavor of the $\Btag$  can be inferred by the charge of various particles produced 
in its decay. 
The purest information comes from  high momentum leptons produced in  semileptonic $B$ decays. 
Although less pure than the leptons, charged kaons are also 
an excellent  source of tagging information 
because they are very commonly produced in $B$ decays. 
Further tagging discrimination can be obtained 
from slow pions produced in the decay of a $D^{\pm *}$, 
from very energetic pions produced from low multiplicity hadronization of the $W$ boson in the 
decay $b\to cW$,  
or from $\Lambda$ hyperons. 

The tagging algorithm combines the above  information into 
a discriminating variable that takes into account correlations 
between different sources of tagging information. This is achieved in 
BaBar by the use of an artificial neural network and in Belle 
by the use of a likelihood function implemented in a look-up table.
The BaBar (Belle) tagging algorithm  
assigns each event to one of four (six) hierarchical and mutually exclusive tagging categories 
based on the purity of the tag.  

The performance of the tagging algorithms is measured by two quantities:
\begin{itemize} 
\item tagging efficiency ($\epstag$) 
      defined as the probability of assigning a \Bz\ or \Bzb\ tag, 
\item mistag fraction ($w$) defined as the fraction of wrong tags present in the tagged sample.
\end{itemize} 
Tagging purities and efficiencies are used to calculate  the so called 
``effective tagging efficiency'' ($\epseff$), defined as 
\[
	 \epseff = \epstag(1-2w)^2.  
\]
The sensitivity to $A_{CP}(t)$ depends on $1/\sqrt{\epseff}$ to  first approximation, so 
 good performance of the tagging algorithm is crucial to the accuracy of the measurement. 
More importantly,
the observed $CP$ asymmetry is related to that defined in (\ref{acpt}) 
by the expression  $A_{CP}^{\mathrm{obs}}(t) = (1-2w) A_{CP}(t)$. 
Thus, an inaccurate determination of the wrong tag fraction $w$ 
would bias the measured amplitude of the $CP$ asymmetry. 
For this reason, the tagging parameters are measured directly from data 
in the context of the time dependent 
\Bz\ mixing analysis. 
In this analysis, one of the two \Bz\ mesons produced in the $\Upsilon(4S)$ decay 
is reconstructed into a  flavor eigenstate ($\Bflav$) 
while the other 
is tagged using the algorithm described above.  
The flavor eigenstates considered for this analysis include hadronic decays such as  
$\Bz\to J/\psi K^{*0}$($K^{*+}\to K^+\pi^-$)  and  $\Bz\to D^{-(*)}h^+$  
with $h^+=\pi^+$, $\rho^+$ or $a_1^+$ and semileptonic 
decays such as $\Bz\to D^{*-}l^+{\nu}$. 

The mistag fraction ($w$) can be extracted from the fit of the time 
dependent mixing asymmetry, $\Amix(t)$, defined as: 
\[
	\Amix(t) = \frac{\Nunmixed(t)-\Nmixed(t)}{\Nunmixed(t)+\Nmixed(t)} 
	=  (1-2w) \cos(\Delta m t), 
\]
where $\Nmixed$ is the number of events with two \Bz\ or two \Bzb\ tags, while 
$\Nunmixed$ is the number of events with one \Bz\ and one \Bzb\ tags. 
The result of the fit to $\Amix(t)$ is shown in figure \ref{figMixing}: 
the distribution on the left refers to the BaBar measurement of $\Dm$ 
using fully reconstructed hadronic final states\cite{mixing},  
while the distribution on the right refers to the Belle measurement of $\Dm$ 
using semileptonic decays\cite{mixing2}.  
The amplitude of the mixing asymmetry measures directly the ``tagging dilution'' $(1-2w)$. 
\begin{figure}[t]
\begin{center}
\psfig{file=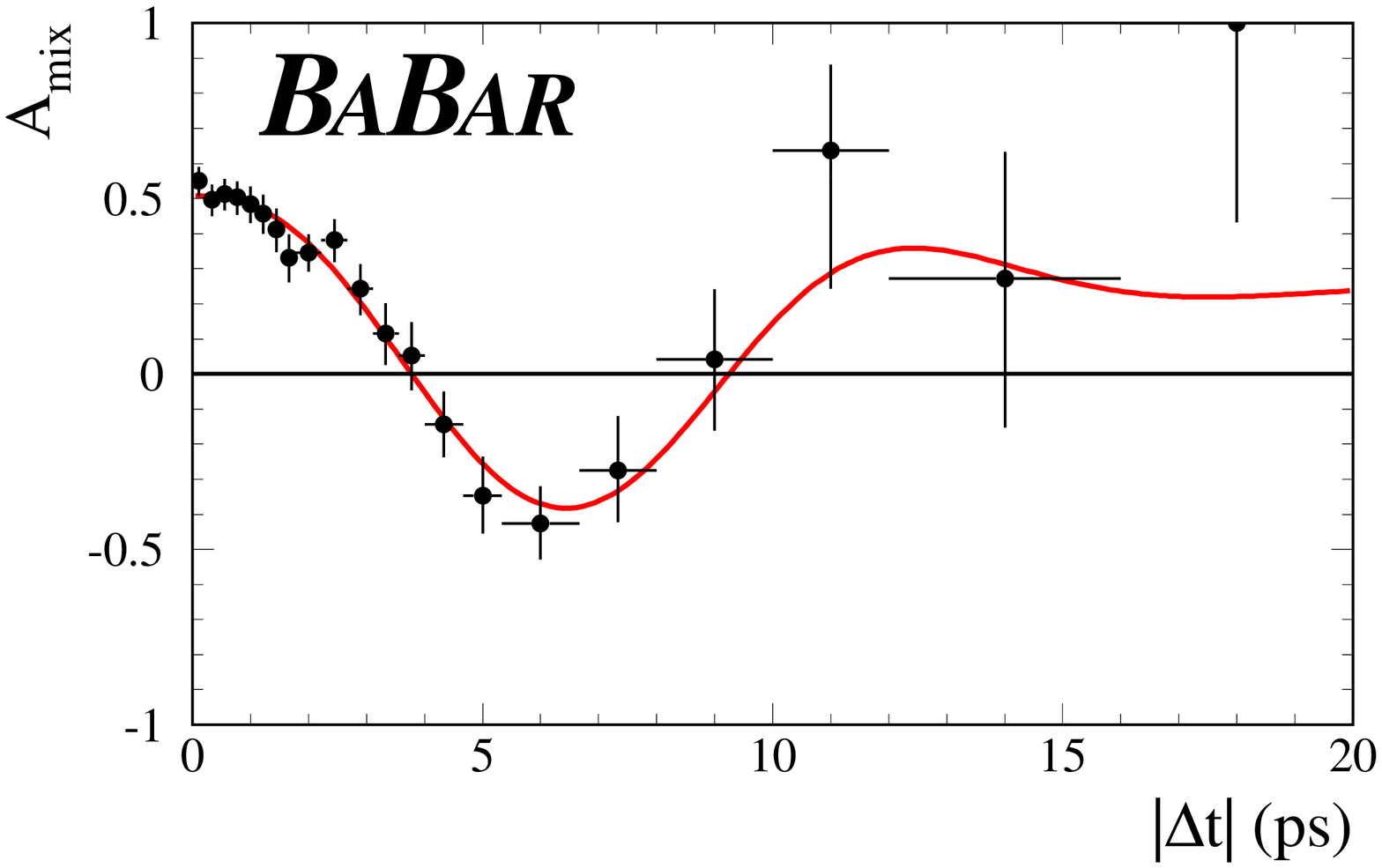,width=2.4in,height=1.5in}
\psfig{file=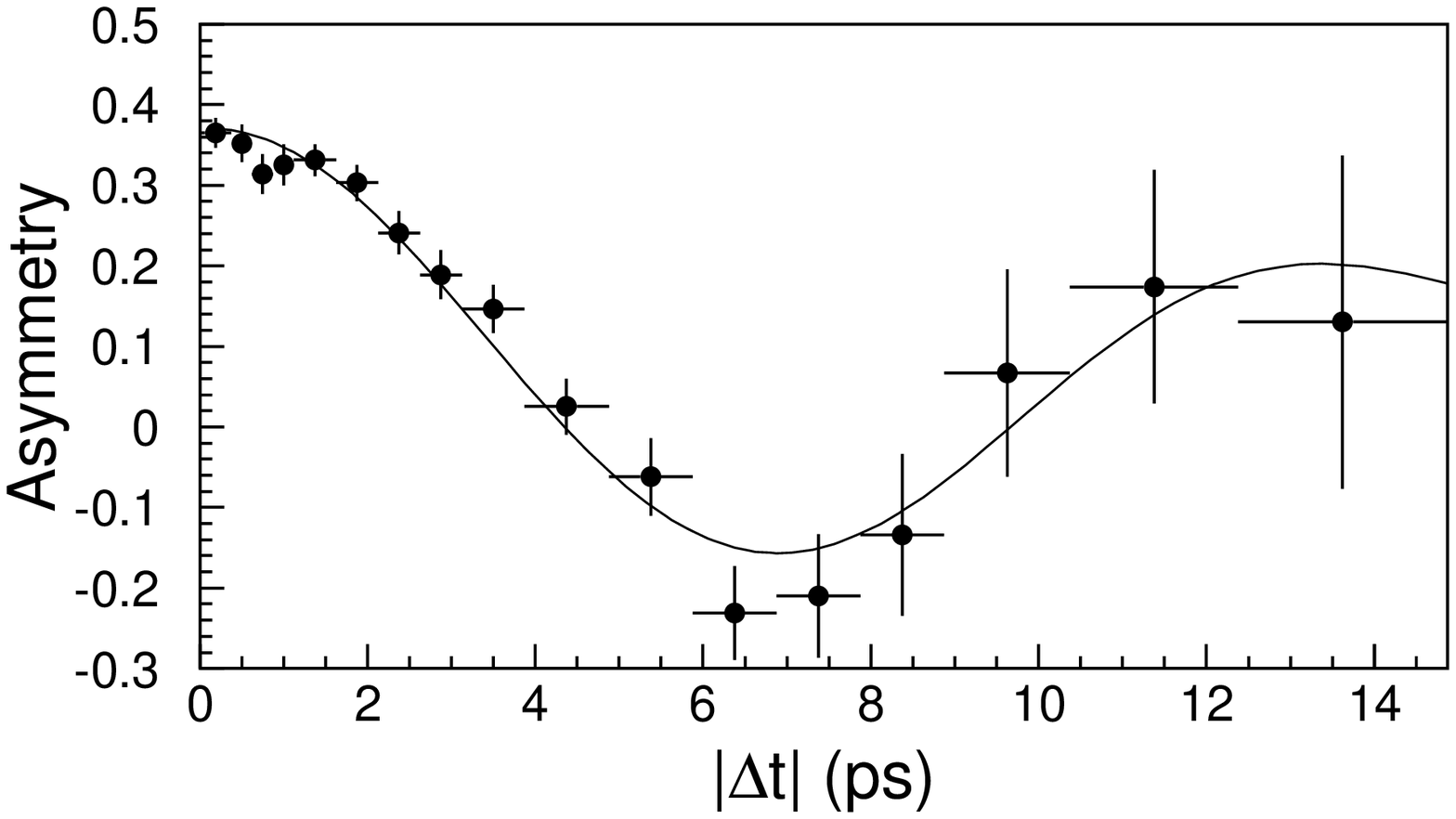,width=2.5in,height=1.75in}
\end{center}
\caption{Time-dependent mixing asymmetry for decays of neutral $B$ into flavor eigenstates with, 
superimposed, the projection of the unbinned maximum likelihood fit. 
The distribution on the left refers to the BaBar measurement 
using fully reconstructed hadronic final states,  
while the distribution on the right refers to the Belle measurement 
using semileptonic decays.} 
\label{figMixing}
\end{figure}

The overall effective tagging efficiency is measured to be
$\epsilon_{\mathrm{eff}}=(28.1\pm0.7)\%$ 
in BaBar and  $\epsilon_{\mathrm{eff}}=(28.8\pm0.6)\%$ in Belle.

\subsection{$\Delta t$ determination} 

The difference in decay times of the two $B$  mesons (\Dt )
can be inferred from the measured distance between 
the two decay vertices ($\Delta z$). 
Since both $B$ mesons are produced with a known boost 
 parallel to the collision ($z$) axis, the time between the 
decays is given by $\Delta t=\Delta z/(\beta \gamma c)$. 
The average separation between the two $B$ decay vertices is  
about 250 $\mu$m in BaBar and 200 $\mu$m in Belle. 

The position of the decay vertex of the \Bz\ decaying into a 
$CP$ eigenstate ($B_{CP}$) is easily reconstructed by fitting 
the charged tracks that appear in the decay to a common vertex. 
The resolution obtained for this vertex is about 60 $\mu$m. 

The reconstruction of the decay vertex  of the other $B$ meson ($\Btag$) is more difficult 
because it is not possible to completely separate tracks originating promptly from the $B$ decay from those 
originating from decays of $D$ mesons produced in $B \to DX$. Because of the relatively long lifetime of the
$D$ mesons, the latter tracks produce a bias when used in the reconstruction of the $\Btag$ decay vertex. 
The position of this vertex is obtained by a vertexing algorithm that uses all 
the charged tracks that do not belong to the $CP$ side along with constraints from the beam 
spot location. Tracks identified as decay products of  $K^0_S$, $\Lambda$ or photon conversions,  
as well as poorly reconstructed tracks, are excluded from the vertex. 
To minimize the bias due to  $D$ meson decay products, the algorithm also 
removes those tracks that give large contribution to the overall $\chi^2$ of the fit. 

Since the resolution on $\Delta t$ is totally dominated by the tagging side, 
it is independent from the decay mode of the $B_{CP}$. 
This allows us to determine the parameters of the resolution function from an 
independent sample of $B$ mesons exclusively reconstructed into flavor eigenstates. 
The time resolution is measured to be about 1.1 ps for BaBar and 1.4 ps for Belle.

\subsection{CP asymmetry fit } 

The $CP$ asymmetry amplitudes are determined from an unbinned maximum likelihood fit 
to the $\Delta t$ distributions for events tagged as \Bz\ and \Bzb .
The BaBar analysis\cite{babarPRD} uses a combined fit to determine simultaneously the $CP$,  
tagging and vertexing parameters from the $B_{CP}$ and $\Bflav$ samples. 
This approach takes better  account of the correlations between the various parameters. 
Belle\cite{bellePRD}, instead, prefers to disentangle the fit of the $CP$ 
asymmetries from the determination of the vertexing and tagging parameters, 
thereby simplifying the final fit and the study of systematics.

\section{The ``golden'' measurement of the angle $\beta$  }	

The decays $\Bz\to\text{charmonium}+K^0$ are known as the ``golden modes'' for the measurement of the angle 
$\beta$ of the UT. These decays are dominated by a tree level diagram $b\to c\overline{c}s$ with 
internal $W$ boson emission. 
The leading penguin diagram contribution to the final state has the same weak phase as the tree diagram, 
and the largest term with different weak phase is a penguin diagram contribution suppressed by $O(\lambda^2)$. 
This makes $|\lambda_f|=1$ a very good approximation,
and thus $A_{CP}(t) = \text{Im}(\lambda_f)\sin(\Dm\Dt)$. 

For the ``golden modes'', $\lambda_f$ is given by: 
\begin{equation}
\lambda_f = \eta_f \left (\frac{V^*_{tb}V_{td}}{V^*_{td}V_{tb}} \right )
\left (\frac{V^*_{cs}V_{cb}}{V^*_{cb}V_{cs}} \right )
\left (\frac{V^*_{cd}V_{cs}}{V^*_{cs}V_{cd}} \right )
= \eta_f \ e^{-2i\beta}
\label{lambdaPsi}
\end{equation}
were the first term comes from \Bz-\Bzb\  mixing, the second from the ratio of the amplitudes 
$\overline{A}_f/A_f$ and the third from $K^0$  mixing. 
The parameter $\eta_f$ is the  $CP$ eigenvalue of the final state, negative 
for charmonium + $K_S$ and positive for  charmonium + $K_L$. 

From (\ref{lambdaPsi}) and (\ref{acpt4}) it follows that 
\begin{equation}
A_{CP}(t) = -\eta_f \sin2\beta \sin(\Dm \Dt)  
\label{acpt5}
\end{equation}
which shows how the angle $\beta$ is directly and simply measured by the amplitude 
of the time dependent $CP$ asymmetry.

Besides the theoretical simplicity, these modes also offer experimental advantages because of  their  
relatively large branching fractions ($\sim 10^{-4}$) 
and the presence of the narrow $J/\psi$ resonance in the final state,
which provides a powerful rejection of combinatorial background.

The $CP$ eigenstates considered for this analysis are $J/\psi K_S$, $\psi$(2S)$K_S$, 
$\chi_{c1}K_S$, $\eta_cK_S$, $J/\psi K_L$  and 
$J/\psi K^{*0}$($K^{*0}\to K_S \pi^0$)\footnote{The final 
state $J/\psi K^{*0}$ does not have a definite $CP$, but its 
$CP$ even and odd components can be disentangled using an angular analysis.}.  
The $J/\psi$  and $\psi(2S)$ are reconstructed from the final states $e^+e^-$  
and $\mu^+\mu^-$; the $\psi(2S)$ is reconstructed also from the  final state $J/\psi\pi^+\pi^-$;  
the $\chi_{c1}$ is reconstructed from the radiative decay $J/\psi \gamma$ and   the 
 $\eta_c$ from the hadronic decays $K_SK^+\pi^-$, $K^-K^+\pi^0$ and $p\overline{p}$. 

Figure \ref{sin2bx} shows the invariant mass distribution for $\Bz\to \text{charmonium} + K_S$
candidates  and the momentum distribution for $\Bz\to J/\psi K_L$ candidates. 
\vspace{-1.0cm}

\begin{figure}[th]
\begin{center}
    \psfig{file=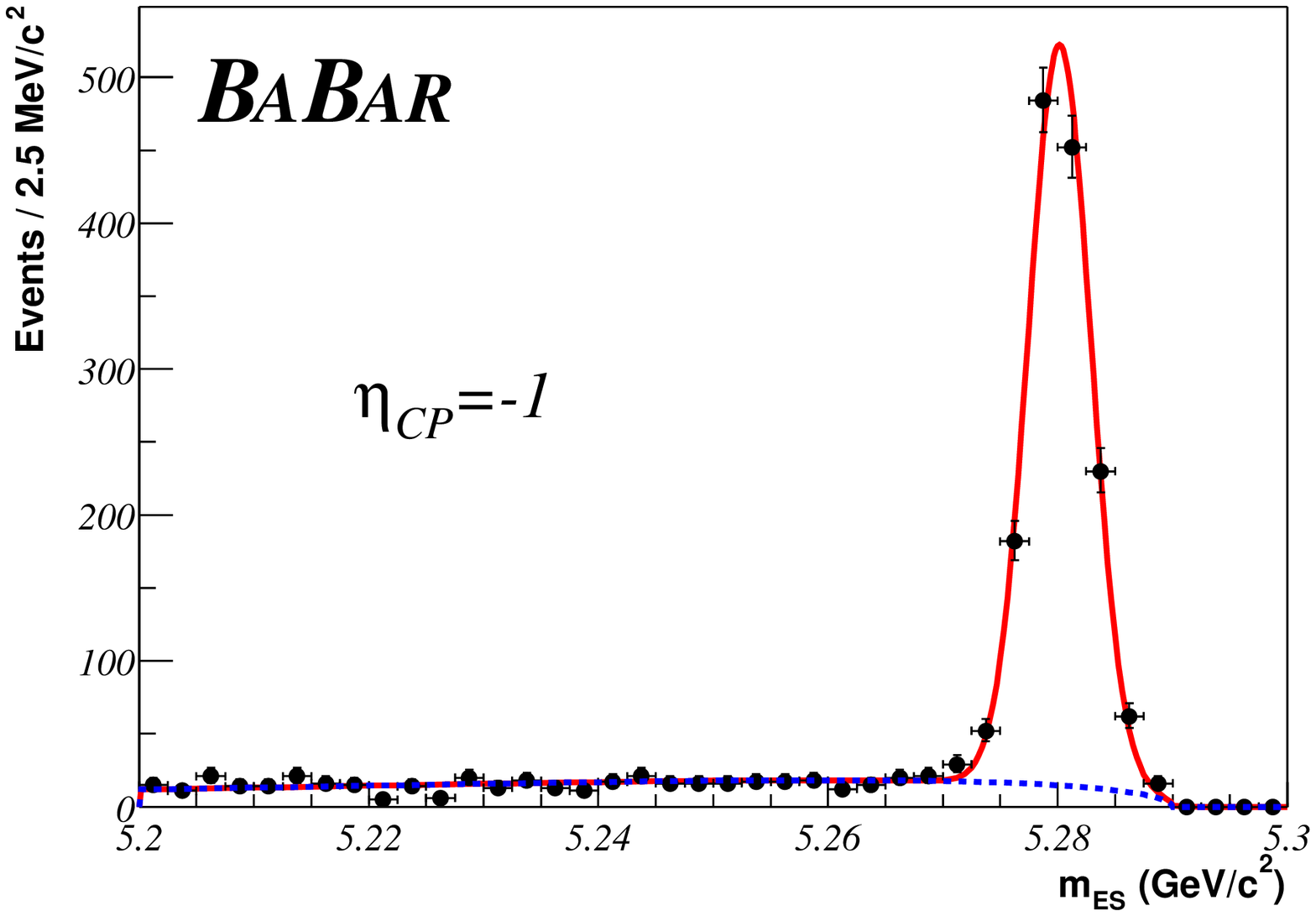,height=2.12in,width=2.2in}
    \psfig{file=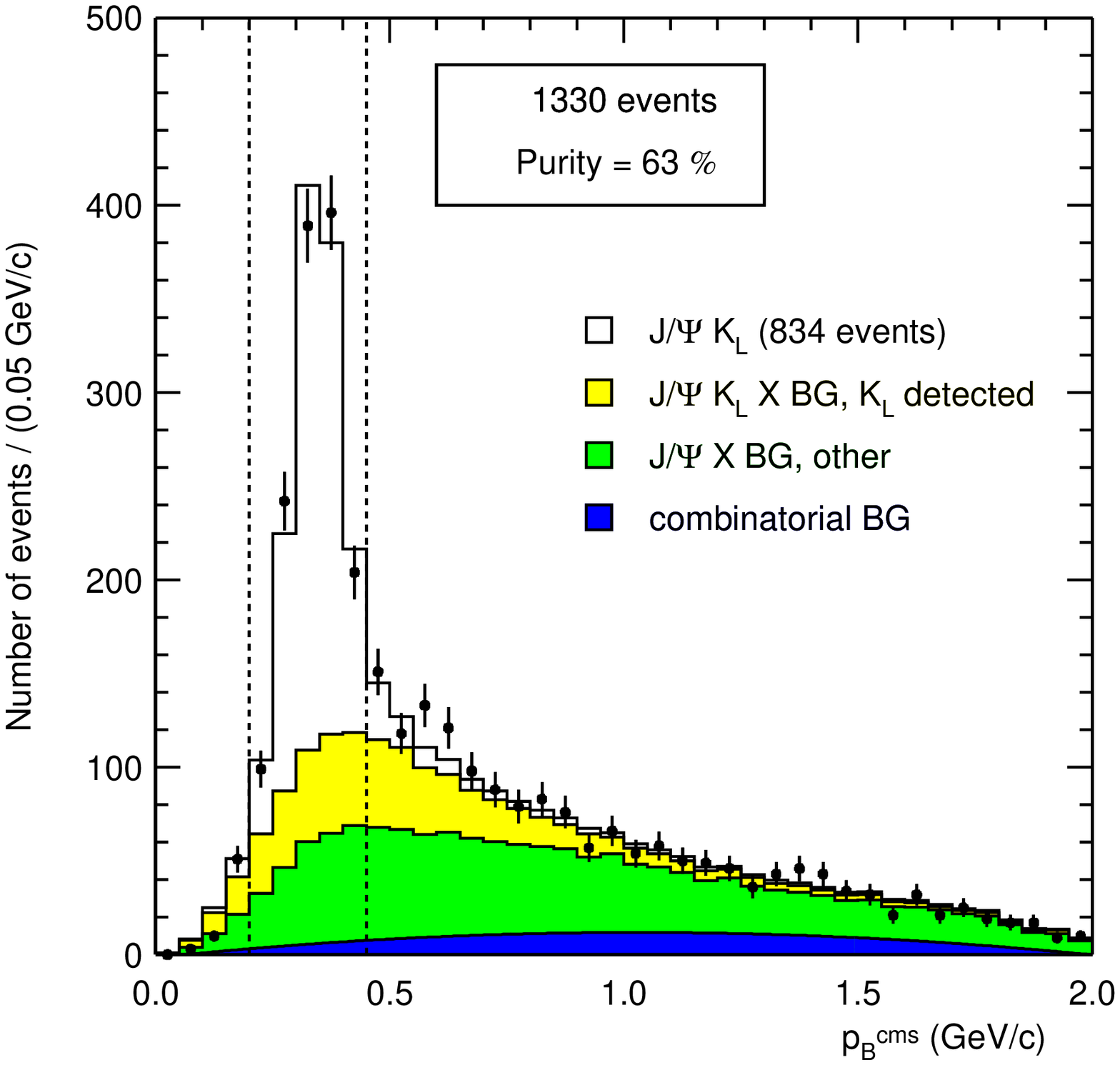,height=2.3in,width=2.4in}
\end{center}
\caption{
Left: invariant mass distribution for the $CP$-odd (charmonium + $K_S$) sample 
as reconstructed by BaBar. The blue curve represent the background contamination. 
Right: distribution of the momentum of the $B$ in the $\Upsilon(4S)$ reference frame for 
$J/\psi + K_L$ candidates as reconstructed by Belle. 
The different background contributions are indicated by the various shades. }
\label{sin2bx}
\end{figure}

The parameter $\sin 2\beta$ is determined by fitting the $\Delta t$ distribution separately 
for events tagged as \Bz\  and \Bzb . 
The asymmetry between the two \Dt\ distributions, clearly visible in figures \ref{babarsin2b-2} (BaBar) and 
\ref{belles2b-3-4} (Belle), 
is a manifestation of $CP$ violation in the $B$ system. 
The same figures also display the corresponding raw $CP$ asymmetry with the 
projection of the unbinned maximum likelihood fit superimposed. 
\begin{figure}[th]
\begin{center}
\includegraphics[width=0.5\linewidth]{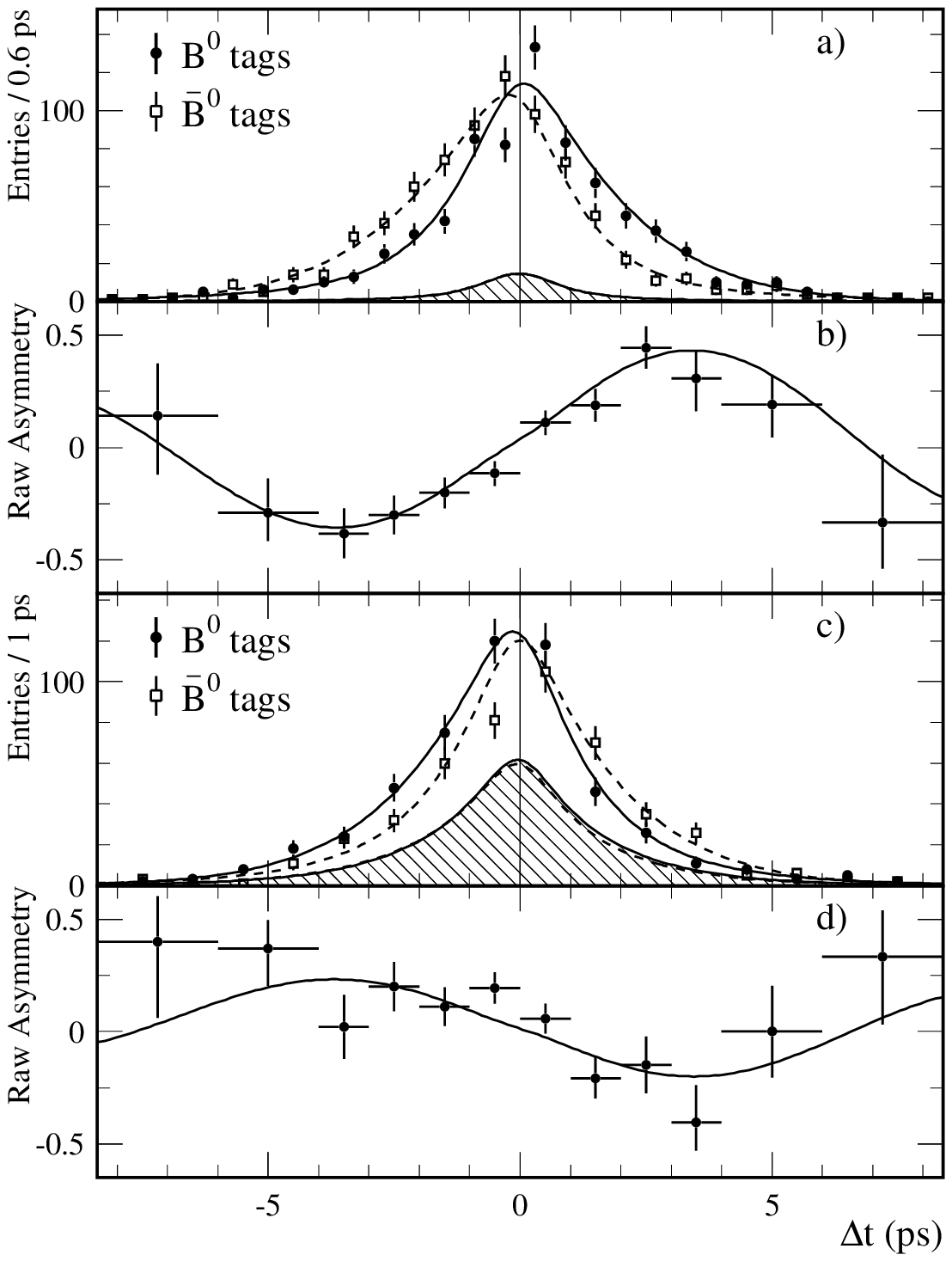}
\end{center}
\caption{Measurement of $\sin 2\beta$ in the ``golden modes'' by the BaBar Collaboration.  
Figure a) shows the \Dt\ distributions for events tagged as \Bz\ (full dots) or \Bzb\ (open squared) 
in $CP$ odd  (charmonium $K_S$) final states. Figure b) shows the  
corresponding raw $CP$ asymmetry with, superimposed, the projection of the unbinned maximum likelihood fit. 
Figure c) and d) contain  the corresponding information for $CP$ even ($J/\psi K_L$) final states. }
\label{babarsin2b-2}
\end{figure}
\begin{figure}[th]
  \begin{center}
    \psfig{file=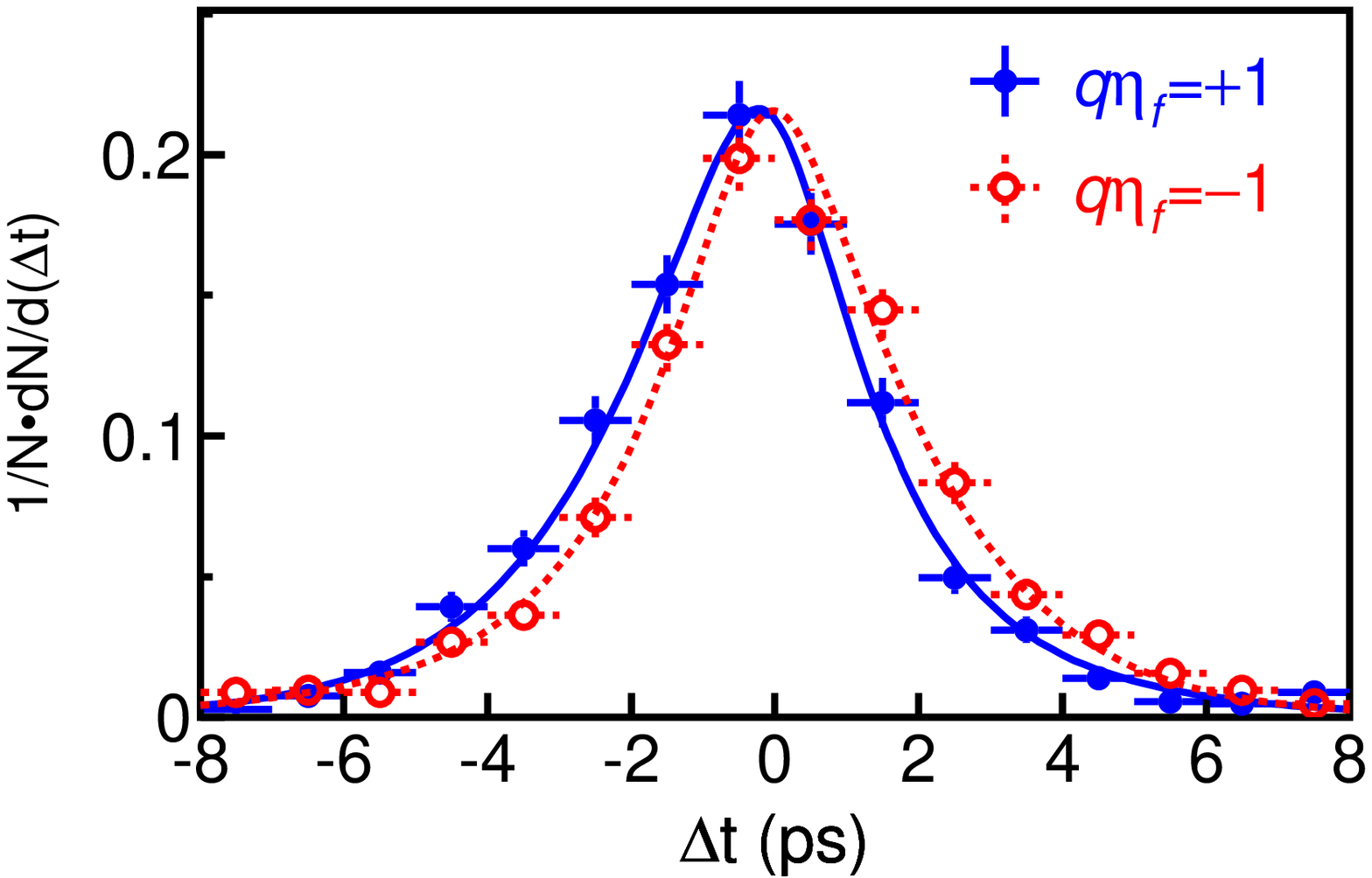,width=2.3in,height=2.0in}
    \psfig{file=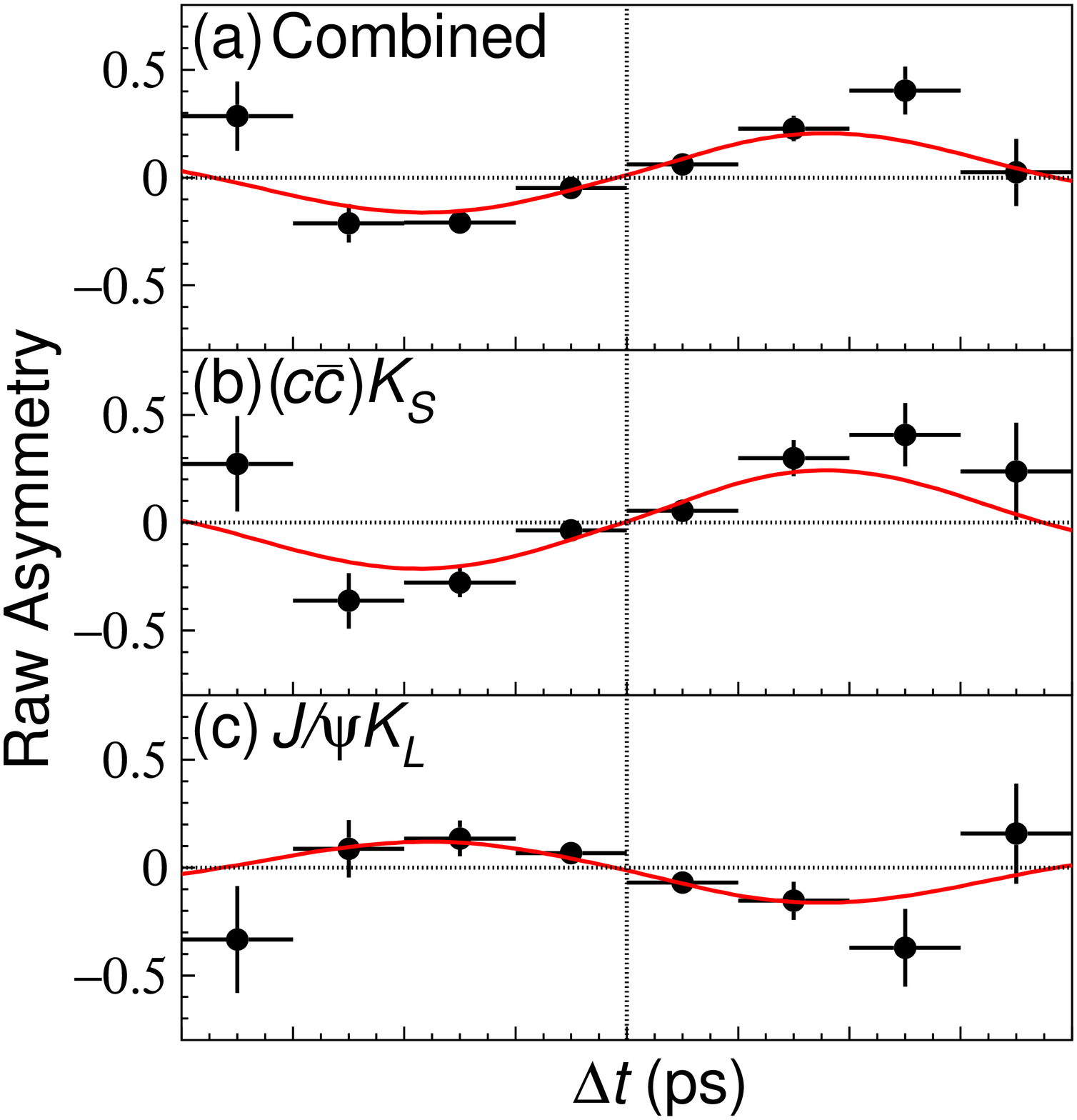,width=2.5in,height=2.0in}
  \end{center}
\caption{Measurement of $\sin 2\beta $ in the ``golden modes'' for the Belle Collaboration.  
  The figure on the left shows the \Dt\ distributions for events with $q\eta_f=+1$ (full dots) and 
$q\eta_f=-1$ (open dots), where $q=+1 (-1)$ for events tagged as \Bz (\Bzb ).  
  The figure on the right  shows the  corresponding raw $CP$ asymmetry with, 
  superimposed, the projection of the unbinned maximum likelihood fit for the full sample (a) and 
  separately for the $CP$ odd (b) and $CP$ even (c) samples.} 
\label{belles2b-3-4}
\end{figure}

The results of the fits are $\sin 2\beta=0.741\pm 0.067\pm 0.034$ for BaBar\cite{babarsin2b}
and $\sin 2\beta=0.719\pm 0.074\pm 0.035$  for Belle\cite{bellesin2b}. 
The main sources of systematic errors are uncertainties in the background level and characteristics, 
in the parameterization of the \Dt\ resolution,  
and in the measurement of the mistag fractions. Most of these uncertainties 
will decrease with additional statistics, and the systematic error is not expected to dominate 
this measurement at the existing $B$ factories in the foreseeable future. 

The world average value for  $\sin 2\beta$, heavily dominated by the $B$ factory results described above, 
is $\sin 2\beta = 0.734\pm 0.055$. 
This value can be compared with the indirect constraints on the apex of the UT originating from 
measurements of $\epsilon_K$, $|V_{ub}|$, $|V_{cb}|$, \Bz\ and $B_S$ mixing 
as described in reference~\cite{CKMFitter}. 
The comparison, illustrated in figure \ref{CKMFitter}, shows excellent agreement between the 
measurements, indicating  that the observed $CP$ asymmetry is consistent with the 
CKM mechanism being the dominant source of $CP$ violation in flavor changing processes at low energies. 
\begin{figure}[th]
\begin{center}
\includegraphics[width=0.7\linewidth]{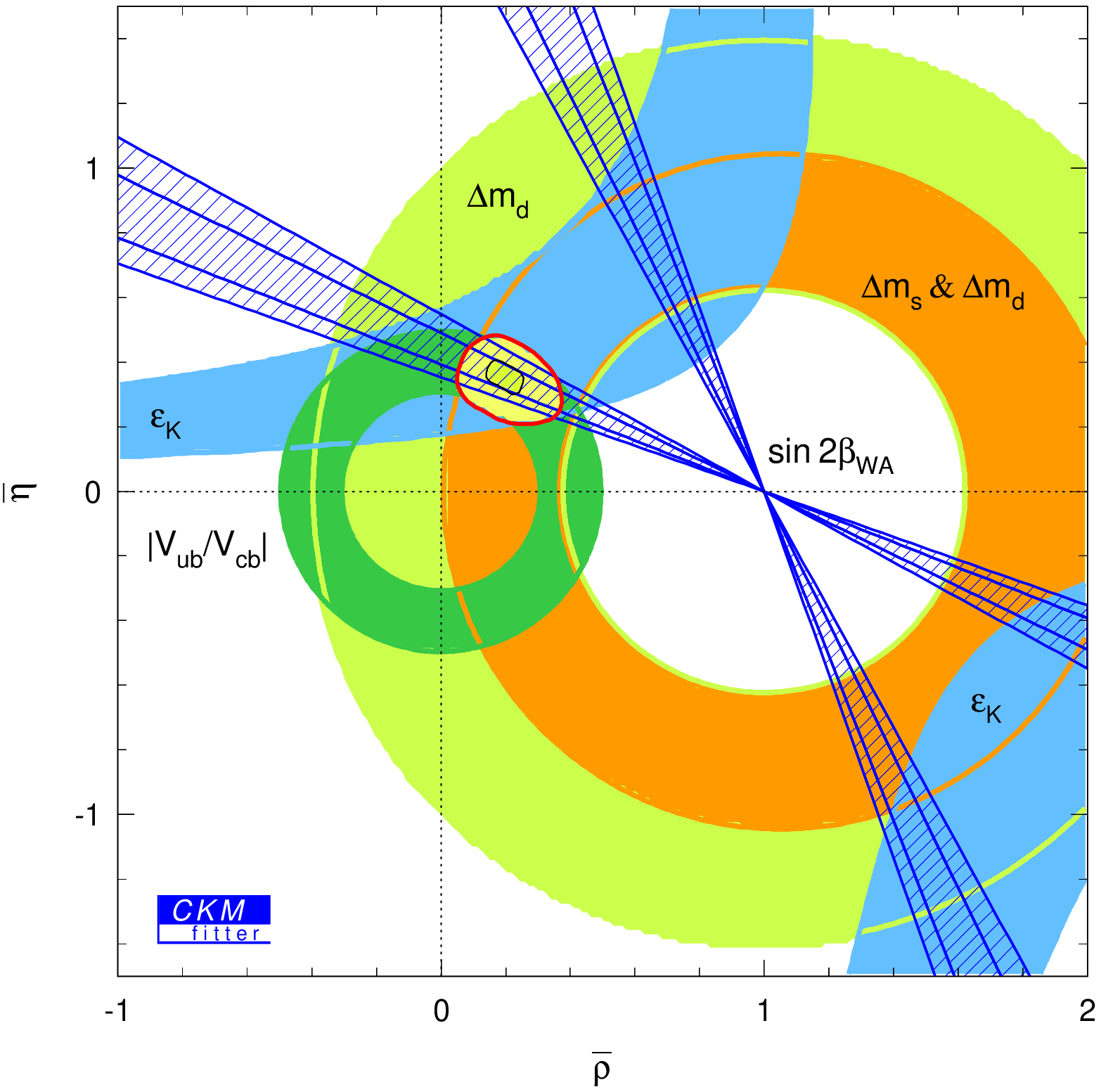}
\end{center}
\caption{Confidence levels in the ($\rho,\eta$) plane obtained from the CKMfitter program. 
         The constraint from the world average $\sin 2\beta$ from $B$\to charmonium + $K^0$ is 
         overlaid in blue with the 1 and 2 $\sigma$ error bands. }
\label{CKMFitter}
\end{figure}

\section{$CP$ violation as a probe for New Physics   }	

$CP$ violation is an excellent probe for seeking  New Physics. 
The simplicity of the CKM mechanism, with a single source of $CP$ violation,
allows for testable predictions. 
The cleanliness of the predictions 
is what makes these studies sensitive to effects of physics beyond the Standard Model that  
may introduce additional sources of $CP$ violation, 
in discrepancy with the predictions of the CKM mechanism.   

Despite the excellent agreement between the measurement of $\sin 2\beta$ 
in decays of $\Bz\to\text{charmonium}+K^0$ and the constraints on the apex of the UT, 
New Physics is not ruled out yet\cite{NirICHEP}. 
In fact, it may manifest itself  in $CP$ violation in other final states such as  
$\phi K_S$,  $\eta 'K_S$, $K^+K^-K_S$, $D^{*+}D^{*-} $ and $D^{*\pm}D^{\mp}$. 

The decays $\Bz\to \phi K_S$ and  $\Bz\to \eta^{\prime}K_S$ are particularly suited for these studies. 
In the SM, these decays are dominated by penguin diagrams; a tree level diagram could contribute to 
$\Bz\to \eta^{\prime}K_S$ but its contribution is small because this decay is both Cabibbo and color 
suppressed. If, as expected\cite{Grossman}, $|\lambda_f|\sim 1$, then the amplitude 
of the time dependent $CP$ asymmetry in these decays would measure the parameter 
$\sin 2\beta$ in a theoretically clean way.   
New Physics could alter these expectations 
via, for example, gluonic penguin diagrams with intermediate squarks 
and gluinos\cite{NirICHEP}. 

BaBar\cite{babarPenguins} and Belle\cite{bellePenguins}  have measured $CP$ violation 
in $\Bz\to \phi K_S$ and  $\Bz\to\eta^{\prime}K_S$  decays, and their results are summarized in figure 
\ref{tabPenguins}. 
Although these measurements still suffer from large statistical uncertainties, 
it is tempting to compare the parameter $\sin 2\beta$ measured in the penguin-dominated
modes with the same parameter measured in the ``golden modes''. 
The comparison shows a $2.7\sigma$ discrepancy for the theoretically cleaner channel 
$\Bz\to\phi K_S$ and $1.6\sigma$ discrepancy for $\Bz\to\eta^{\prime}K_S$. 
Although it is premature to interpret these measurements   as a hint of New Physics, these results 
are certainly intriguing and they have generated much interest  in the  community\cite{theoPenguins}.

\begin{figure}[th]
\begin{center}
\includegraphics[width=3in]{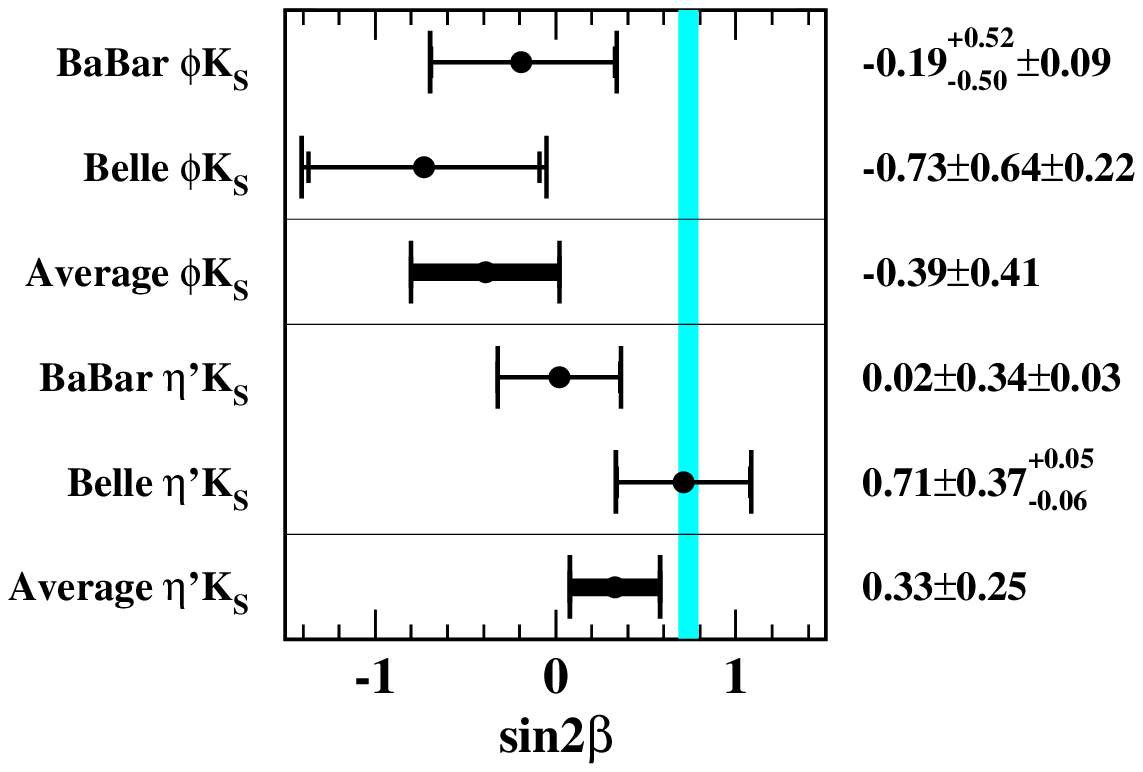}
\end{center}
\vspace{-0.30cm}
\caption{Summary and averages of the measurements of ``$\sin 2\beta$'' in
$\Bz\to\phi K_S$ 
and  $\Bz\to\eta 'K_S$. The vertical shadowed region is the world average measurement of $\sin2\beta$ 
in the ``golden mode'' $\pm 1 \sigma$.}
\label{tabPenguins}
\end{figure}

\section{The measurement of the angle $\alpha$}	

If the decay $\Bz\to\pi^+\pi^-$  were dominated by the $b\to u$ tree level diagram, 
the amplitude of the time dependent 
$CP$ asymmetry in this channel would be a clean measurement of the parameter $\sin 2\alpha$. 
Unfortunately, the contribution of gluonic $b\to d$ penguin amplitudes to this 
final state cannot be neglected. 
The ratio between the penguin and tree contributions can be estimated 
from the ratio BF(\Bz\to $K\pi$)/BF(\Bz\to $\pi\pi$) to be $|P/T|\sim 30\%$. 
These contributions have a different weak phase and additional strong phases\cite{PiPipenguins}.  
As a result, in the study of time dependent $CP$ asymmetry  
one has to fit for both the sine and the  cosine terms in equation (\ref{acpt2}). 
The coefficient of the sine term $S_{\pi\pi}$ can be related to the 
angle $\alpha$ through isospin symmetry\cite{GronauLondon}, while 
the coefficient of the cosine term  $C_{\pi\pi}$  measures direct $CP$ violation. 

Experimentally, this analysis is very challenging for several reasons. First, the final 
state of interest, $\Bz\to\pi^+\pi^-$, has to be disentangled from the similar decay 
$\Bz\to K^+\pi^-$, which has a much  larger branching fraction.
The excellent particle identification 
capability provided by the Cherenkov detectors of BaBar and Belle  
plays a crucial role in suppressing this background. 

The second experimental challenge is due to the high combinatorial background 
from  \qqbar\ events in which two energetic pions in opposite jets are selected. 
In BaBar\cite{BaBarPiPi}, this so-called ``continuum'' background 
 is suppressed employing a Fisher discriminant 
built with the momentum flow in nine cones around the candidate axis.  
In order to optimize the statistical power of the data sample, no cuts are 
applied on the particle identification and Fisher variables;  
instead they are included in the $CP$  maximum likelihood fit.  

In the Belle analysis\cite{BellePiPi}, the suppression of continuum is instead achieved by the use of 
a likelihood analysis using six modified Fox-Wolfram moments\cite{foxW} and 
the $B$ flight direction. 
The background is rejected by cutting on the output of the likelihood ratio 
$\mathcal{L}_S/(\mathcal{L}_S+\mathcal{L}_B)$ prior to the $CP$ fit, 
where $\mathcal{L}_S$ and $\mathcal{L}_B$ are the likelihood functions for 
signal and background. 

The \Dt\ distribution for events tagged as \Bz\  and \Bzb\  is shown in figures  
\ref{babarpipi} and \ref{bellepipi} for the BaBar and Belle experiments, respectively. 
The bottom plot of these figures displays the corresponding $CP$ asymmetry after 
background subtraction, with the 
projection of the unbinned maximum likelihood fit superimposed.

\begin{figure}[th]
\centerline{\psfig{file=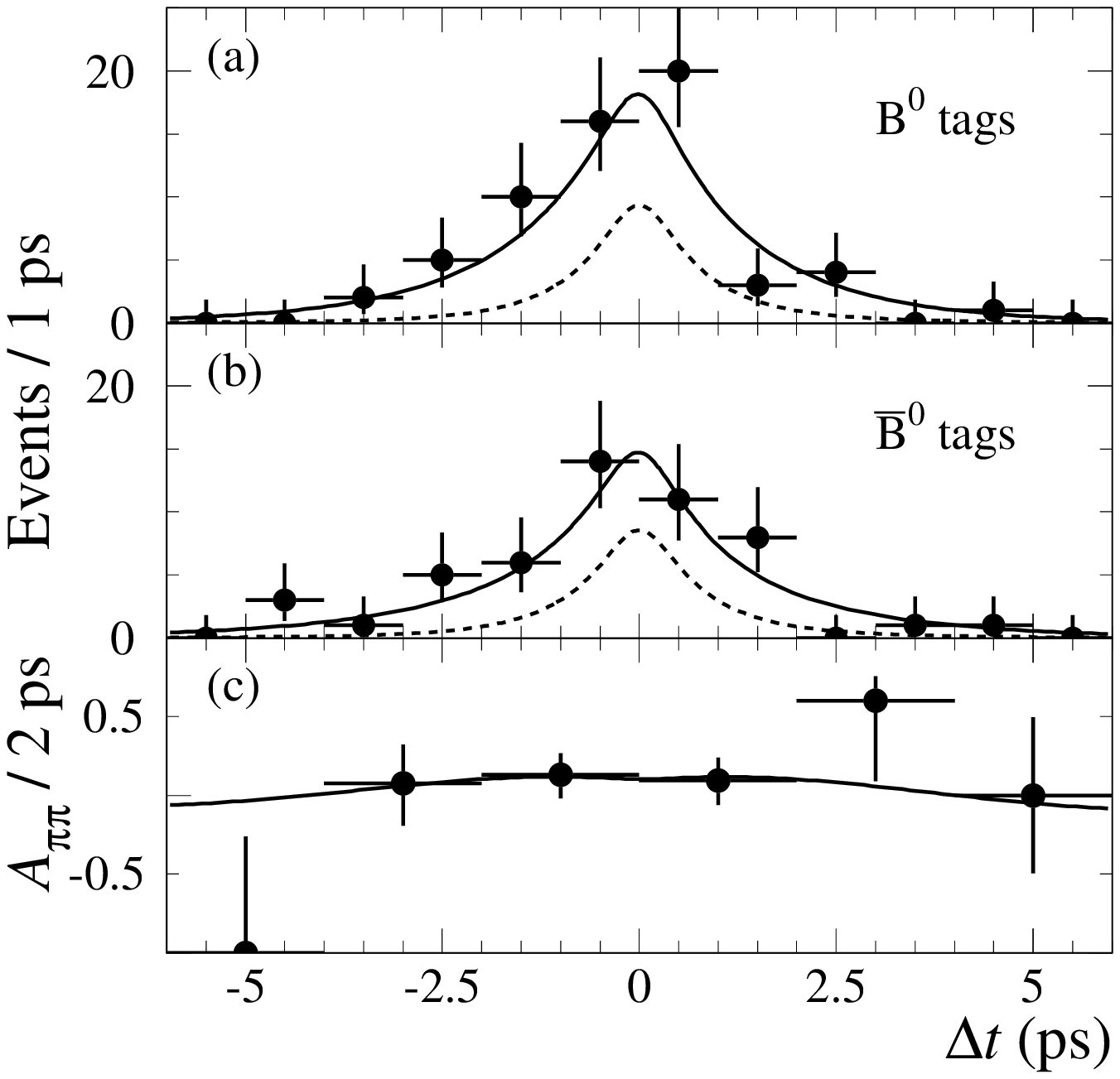,width=1.7in}}
\vspace{0.2cm}
\caption{
Distributions of $\Dt$ as measured by BaBar for events enhanced in 
$\pi^+\pi^-$ decays and tagged as 
 $\Bz$ (a) or  $\Bzb$ (b). The time dependent $CP$ asymmetry
is reported in (c).  Solid curves represent 
projections of the maximum likelihood fit, while dashed curves 
represent the sum of $\qqbar$ and $K\pi$ background events.}
\label{babarpipi}
\end{figure}
\begin{figure}[th]
\centerline{\psfig{file=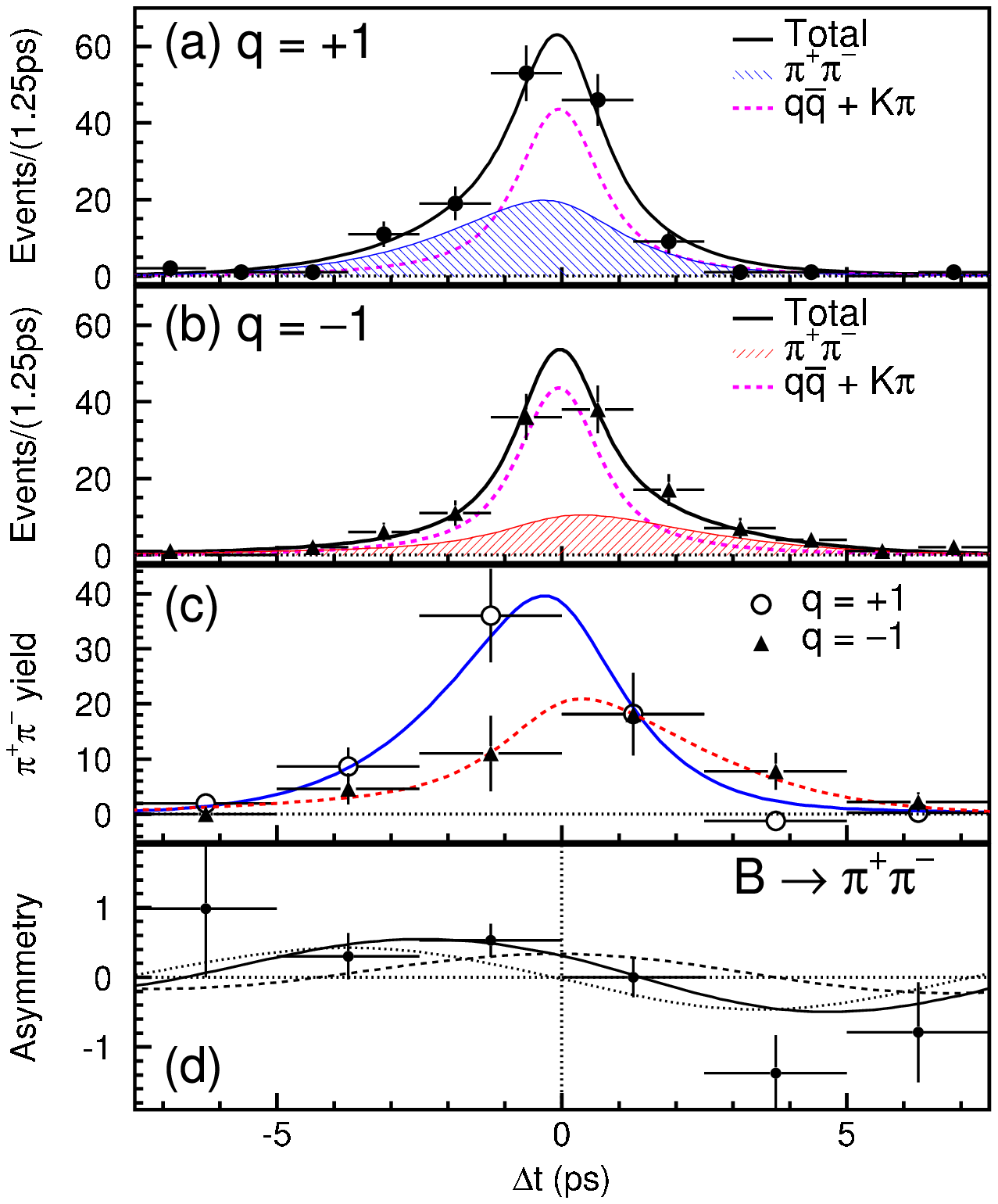,width=2.3in}}
\caption{ 
Distributions of \Dt\ for event enriched in $\pi^+\pi^-$ decays and 
tagged as  $\Bz$ (a) or $\Bzb$ (b) as measured by Belle. 
The background subtracted \Dt\ distributions are shown in (c). 
In (d) is displayed the distribution of the time dependent $CP$ asymmetry 
for the \Bz\to$\pi^+\pi^-$ candidates with,   
superimposed, the projection of the unbinned maximum likelihood fit (solid line). 
The cosine (sine) components are shown as dashed (dotted) lines. 
}
\label{bellepipi}
\end{figure}

The fitted values for the sine and cosine terms of the time dependent $CP$ 
asymmetry obtained in  the BaBar experiment are 
$S_{\pi\pi}={0.02}\pm{0.34}\pm{0.05}$ and 
$C_{\pi\pi}={-0.30}\pm{0.25}\pm{0.04}$, 
both compatible with zero. 
The corresponding results obtained by Belle  are
$S_{\pi\pi}={-1.23}\pm{0.41}^{+0.08}_{-0.07}$  
and 
$C_{\pi\pi}=-0.77\pm 0.27\pm 0.08$, 
indicating that both mixing-induced and direct $CP$ violation effects in charmless $B$ decays are large.  

Since the results from the two experiments are compatible (2.2 $\sigma$ apart), we can average them 
to obtain $S_{\pi\pi}={-0.47}\pm{0.26}$ and 
$C_{\pi\pi}=-0.49\pm 0.19$. 
The average shows evidence for direct $CP$ violation (2.6 $\sigma$), but no compelling 
evidence for mixing-induced $CP$ violation  (1.8 $\sigma$). 
These results can be used to constrain the angle $\alpha$\cite{alpha}. 
The accuracy on the determination of the angle of the UT depends 
critically on the theoretical assumptions used in the fits\cite{alphaerr}. 

A measurement of the angle $\alpha$ can also be extracted from the study of the decay $\Bz\to\rho\pi$. 
Compared with  the 
$\Bz\to\pi\pi$ mode, this final state has the advantage of a 
higher branching fraction  and a smaller penguin pollution. 
On the other hand, this analysis is complicated 
by the presence of four configurations in the final state 
($\Bz\to\rho^{\pm}\pi^{\mp}$, $\Bzb\to\rho^{\mp}\pi^{\pm}$),  
by the fact that  the final state is not a $CP$ eigenstate, 
and by the substantial combinatorial background induced by the 
presence of three pions.  
A theoretically clean extraction of $\alpha$ is possible, but requires a 
combined fit over the entire Dalitz plane. 

The BaBar Collaboration recently published the results of an  
exploratory  $CP$ analysis\cite{RhoPi} that follows a quasi-two-body 
approach and avoids the interference regions in the $\pi^+\pi^-\pi^0$ Dalitz plot. 
The values of the mixing-induced $CP$ violation parameter $S_{\rho\pi}=0.19\pm 0.24\pm 0.03$ 
and 
of the direct $CP$ violation parameter $C_{\rho\pi}=0.36\pm 0.18\pm 0.04$ 
indicate that the accuracy on the extraction of $\alpha$ expected from the Dalitz analysis of 
this decay mode will be  competitive.

\section{Conclusion}	

  Since the beginning of the data taking in 1999, the $B$ factories opened a new chapter in $B$ physics. 
  The excellent performance of the accelerators and detectors allowed the BaBar and Belle Collaborations 
  to accumulate an unprecedented sample of $B$ decays, which lead to  
  the first unambiguous observation of $CP$ violation in the $B$ sector. 

  The measurement of the angle $\beta$ of the Unitarity Triangle allowed the first 
  quantitative test of the the $CP$ sector of the Standard Model. 
  The excellent agreement between direct measurement of the angle $\beta$ and the indirect constraints 
  on the apex of the Unitarity Triangle suggests that the CKM mechanism is the dominant 
  source of $CP$ violation  at low energies. 
    
  Contributions from New Physics could be detected in the measurement of $CP$ asymmetries 
  in penguin dominated  decays, such as $\Bz\to\phi K_S$ or $\eta^{\prime}K_S$. 
  The first measurements of these quantities have been published recently by both Collaborations 
  showing a deviation of $2.7\sigma$ compared to the Standard Model expectations 
  for the $\Bz\to\phi K_S$ channel. 

  The measurement of the angle $\alpha$ of the UT   through a time dependent $CP$ analysis 
  is in progress in the channels $\Bz\to\pi^+\pi^-$ and $\Bz\to\pi^+\pi^-\pi^0$. 
  The results of the $\pi^+\pi^-$ analysis show evidence for direct $CP$ violation. 

  The data analyzed so far by BaBar and Belle corresponds to about 80\,fb$^{-1}$ per experiment. 
  By 2006, each $B$ factory expects to have more than 500\,fb$^{-1}$ available for analysis. 
  The increased data sets will allow a precise test of the $CP$ sector of the Standard Model and 
  will provide a sensitive probe of New Physics.

\section*{Acknowledgments}
It is a pleasure to thank T.~Browder, C.~Dallapiccola, 
D.~Lange, M.~Morii, Y.~Sakai and R.~Yamamoto for their careful reading of the the 
manuscript and  helpful comments.  
This work was supported in part by the DOE contract DE-FC02-94ER40818.

\end{document}